\documentclass[aps,amssymb,pre, amsmath, floatfix, preprint]{revtex4}
\usepackage{graphics}
\usepackage{psfrag}
\usepackage{epsfig}
\usepackage{subfigure}
\begin{document}
\bibliographystyle{prsty}

\title{Long Range Correlations in Granular Shear Flow I:\\  Numerical Evidence}
\author{Gregg Lois$^{(1)}$}
\author{Ana\"el Lema\^{\i}tre$^{(2)}$}
\author{Jean M. Carlson$^{(1)}$}
\affiliation{
$^{(1)}$ Department of physics, University of California, Santa Barbara, California 93106, U.S.A.}
\affiliation{$^{(2)}$ Institut Navier-- LMSGC, 2 all\'ee K\'epler, 77420 Champs-sur-Marne, France} 
\date{\today}

\begin{abstract}
We investigate the emergence of long-range correlations in granular shear flow.  By increasing the density of a simulated granular flow we observe a spontaneous transition from a dilute regime, where interactions are dominated by binary collisions, to a dense regime characterized by large force networks and collective motions.  With increasing density, interacting grains tend to form networks of simultaneous contacts due to the dissipative nature of collisions.  We quantify the size of these networks by measuring correlations between grain forces and find that there are dramatic changes in the statistics of contact forces as the size of the networks increases.  
\end{abstract}
\maketitle

\section{Introduction}
Granular materials exhibit a wide range of fascinating behaviors~\cite{reviewjaeger}, but predictive theories linking the microscopic grain interactions to macroscopic properties remains a topic of much debate.  Of particular interest are constitutive relations for granular flow, which are important for engineering and geophysical applications~\cite{applications,applications2} and challenge the tenets of conventional statistical physics~\cite{reviewleo}.  Because granular materials are athermal their dynamics always occur far from equilibrium and a proper formulation of constitutive relations relies on the construction of non-equilibrium statistical theories that must be sensitive to the interactions between grains. 

Interactions in realistic granular materials arise due to grain elasticity and friction, but are complicated by various other mechanisms including humidity~\cite{humidity0, humidity1,humidity2,humidity3,humidity4}, grain shapes~\cite{angular0,angular1,angular2}, and fracture processes occurring within the material~\cite{fracture0}.  However, a great deal of theoretical and computational progress has been made using the simple approximation that grains are spherical and perfectly dry~\cite{cundallstrack}.  In this case a purely repulsive force arises when two grains 
come into contact due to the deformation of grains and friction between grains.  

Understanding the nature of grain forces and dynamics, even in this relatively simple case, has proven difficult.  At very low densities it can
safely be assumed that only binary interactions occur and constitutive relations can be determined by statistically tracking the repulsive force created in each interaction.  This is the basis of kinetic theory, which has 
been successfully applied to granular flows~\cite{campbellreview, goldhirschreview}.  
However, for very large densities, it is observed that multi-grain contacts always
occur~\cite{kudrolligollubcluster,bonamyexps,blairkudrollicluster} and contact forces are transmitted through ``force chain networks'' formed by the topology of the contact network~\cite{cundallstrack, dantufirst, radjai1, behringerlength}.  For these high densities the forces between contacting grains still arise from grain deformation and friction, but the extent of the interactions is not localized and depends on properties of the force chain networks.

The presence of force chain networks calls into question theories that assume localized interactions and has inspired new models based on properties of the force chains~\cite{qmodel,socolaralpha,claudinstress,nicodemi,socolarforce,ottoforce, bouchaudforce}.  However, although force networks can be visualized, it has proven difficult to measure quantitative correlations between contact forces~\cite{mueth98, lovoll, silbertgrest, frenning}.  This has led to the speculation that force chain networks are simply a perceived correlation, until recently when long-range correlations were measured between the {\em averaged} contact forces in a quasi-static experimental shear flow at high density~\cite{behringerlength}.      

This discovery raises important questions about the proper assumptions to make when constructing theories of granular materials.
For very dilute systems only binary collisions occur and force chain networks do not play a role; for very dense systems force networks are the dominating microscopic interaction.  In order to understand the origin of macroscopic properties in granular flows, it is necessary to pinpoint exactly how and when correlations appear.  

In this paper we measure spatial correlations of the {\em total} force on grains undergoing shear deformation.  This measurement defines a characteristic length-scale $\xi$ quantifying the size of force networks arising from clusters of simultaneously contacting grains.  
We find that $\xi$ grows with packing fraction, diverges at a finite packing fraction, and has measurable effects on the contact forces between grains.  The correlation measurement also provides a natural boundary between dilute flows where only binary collisions occur and dense flows where force networks spontaneously emerge.    

We begin in Section~\ref{regimesection} by briefly outlining the phenomenology of granular shear flow to define the exact regime we will be studying.  In Section~\ref{simsection} we  
discuss the numerical algorithm used to simulate shear flow.  In Section~\ref{xisection} we introduce the spatial force correlation measurement which provides a natural definition for a long-range correlation length and in Section~\ref{forcepdfsection} we show how the size of the correlation length affects the contact forces between grains.  

\section{Granular Shear Flow-- Basic Considerations} 
\label{regimesection}

If no external force is applied to a dry granular material in the absence of gravity, it quickly loses all
its kinetic energy in dissipative collisions, and each grain comes to rest.  If this occurs for a dilute system
there are no residual contacts between any grains and the total energy 
is zero.  However, for granular materials with larger densities, there are contacts between grains in the relaxed state and a non-zero 
residual energy remains due to grain deformation and friction.      

If a shear stress is then applied to the system,
motion only occurs if the stress is large enough to overcome the energy stored in the contacts.  The minimum stress needed to initiate
motion is called the yield stress, which is zero below a critical packing fraction $\nu_c$ and is an increasing function of
packing fraction above $\nu_c$~\cite{aharanovsparks, ohernsilbert, zhangmakse}.  

For granular shear flows with $\nu > \nu_c$, previous research has
demonstrated that the stiffness of the grains plays an important role at all values of the shear rate~\cite{campbellrigid}.
This is because grains are not able to rearrange to a configuration where
no contacts exist and the system moves between different configurations where the grains are always deformed.  
Shear flows with
$\nu>\nu_c$ are characterized by slowly moving quasi-static
flows~\cite{behringerhowell, muethnature}, where force balance is upheld at all times, and jamming~\cite{aharanovsparks, ohernsilbert, zhangmakse}, where motion ceases for stresses below the 
the yield stress.  

Conversely, for granular shear flows with $\nu < \nu_c$, it has been demonstrated that the stiffness of the grains can always be taken large enough so that
it plays no role in the dynamics~\cite{campbellrigid, grestsilbert, dacruz, comparemdcd}.  This is because grains are always able to rearrange to find free volume and the system moves between different configurations with minute grain deformation.  In this regime inertial terms are dominant and an invariance in Newton's equations~\cite{Lemaitreprl} shows that the dynamics are controlled exclusively by
the shear rate $\dot\gamma$~\cite{gaj1}.  For these inertial flows the stress tensor 
is proportional to $\dot\gamma^2$, which is referred to as Bagnold's scaling and has been observed in experiments~\cite{bagnold54} and simulations~\cite{grestsilbert, dacruz, gaj1}.

In this paper we consider only inertial flows of granular materials with $\nu< \nu_\mathrm{c}$, in the limit of infinite stiffness.  This allows us to investigate a wide range of packing fraction, from zero to $\nu_\mathrm{c}$, and explore how correlations between grains naturally emerge in this range.  In particular, we measure spatial force correlations and observe that they decay exponentially with a characteristic length-scale $\xi$.  By measuring distributions of contact forces, we demonstrate that only binary collisions are relevant for small $\xi$ and simultaneous contacts have quantitative effects on contact forces for large $\xi$.  This allows us to define an important packing fraction $\nu_\mathrm{bc}$, where the crossover occurs between binary collisions and force networks.  This crossover is accompanied by a signature in the contact force probability distribution function that has been observed previously in experiments on hopper flow~\cite{hopperexps}.

\begin{figure}
\psfrag{a}{\Huge{$\nu_\mathrm{bc}$}}
\psfrag{b}{\Huge{$\nu_\mathrm{c}$}}
\psfrag{c}{\Huge{$\nu$}}
\psfrag{flows,}{$\mathrm{Flows}$}
\resizebox{!}{.38\textwidth}{{\includegraphics{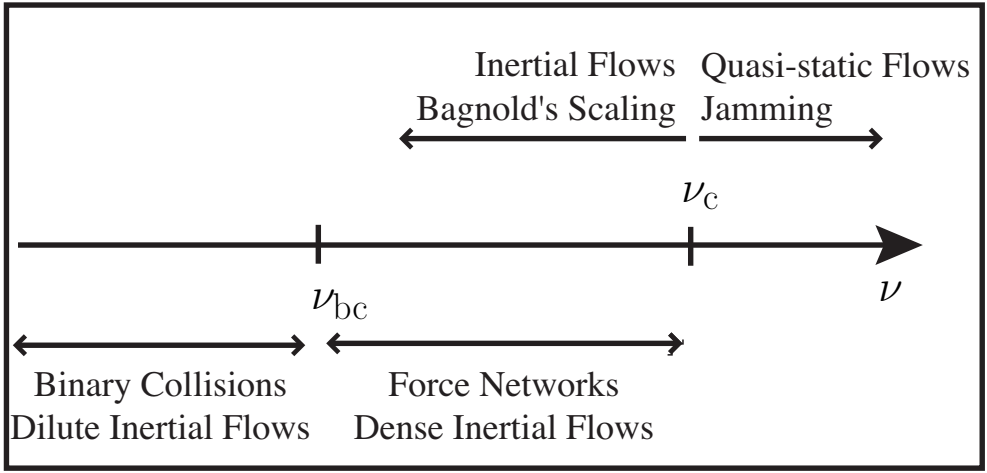}}}
\caption{ \label{phasediagram0} Schematic of the regimes of granular shear flow.  Here we explore inertial flows with $\nu<\nu_\mathrm{c}$ and show that there is a network transition at $\nu_\mathrm{bc}$ where force chain networks begin to form and affect the dynamics.}
\end{figure}

The transition at $\nu_\mathrm{bc}$ is purely dynamical and would not occur if $\dot\gamma=0$.  We find that the value of $\nu_\mathrm{bc}$ depends on the amount of energy dissipated at each contact, but it is always the case that $\nu_\mathrm{bc}$ is strictly less than $\nu_\mathrm{c}$.  This separates the inertial regime into a dilute inertial regime where only binary collisions are relevant and a dense inertial regime where clusters of contacting grains form into complex force networks.  The network transition at $\nu_\mathrm{bc}$ arises due to steric exclusion and is independent of the roughness of the grains.  A schematic of the different regimes available to a granular shear flow is displayed in Figure~\ref{phasediagram0}.

\section{Simulations}
\label{simsection}
We perform simulations of inertial granular shear flow in two-dimensions using the Contact Dynamics (CD) algorithm.  The CD algorithm 
was developed to model the dynamics of dense collections of rigid grains~\cite{CD1,CD2,CD3,CD4,CD5,CD6} and can include the effects of long
lasting, simultaneous contacts between groups of grains.  It is a fully dynamical and athermal algorithm, with the motion of each grain determined at every
time step by integrating Newton's equations.

To carry out the integration, the algorithm provides values for the forces between each pair of contacting grains.
In dry granular materials, contact forces arise due to deformation of the grains upon contact and friction between grains.  The CD algorithm considers the rigid limit in a self-consistent way where forces take on the precise value necessary to prevent deformation, uphold Coulomb friction, and adhere to Newton's equations~\cite{CD1, CD2, CD3, CD4}.   If we use $\mbox{\boldmath $\hat{\sigma}$}^{ij}$ to
denote the unit vector connecting the centers of two contacting grains labeled $i$ and $j$, then the deformation produces a normal force in the direction
of $\mbox{\boldmath $\hat{\sigma}$}^{ij}$ and the interplay of friction and deformation produces a tangential force perpendicular to $\mbox{\boldmath $\hat{\sigma}$}^{ij}$.  These contact forces are 
dissipative and depend sensitively on the velocities of the colliding grains~\cite{lougebin}. 

The CD algorithm determines the forces arising from grain deformation by assuming that grains are infinitely rigid and setting constraints on the total energy dissipated in each contact.  For two grains in contact with a relative velocity ${\bf v'}^{ij}$, the algorithm determines a contact force such that the relative velocity in the next time step ${\bf v}^{ij}$ is given by
\begin{equation}
{\bf v}^{ij} \cdot \mbox{\boldmath $\hat{\sigma}$}^{ij} = -e {\bf v'}^{ij} \cdot \mbox{\boldmath $\hat{\sigma}$}^{ij} \quad ; \quad {\bf v}^{ij} \times \mbox{\boldmath $\hat{\sigma}$}^{ij} = e_t {\bf v'}^{ij} \times \mbox{\boldmath $\hat{\sigma}$}^{ij}.
\label{dynamicalrule}
\end{equation}
In this way, the relative velocities are altered by restitution coefficients in the normal direction ($e$) and tangential direction ($e_t$).     
This procedure ensures that the momentum transfered between grains with non-zero relative velocities is at least that which is supposed to arise from collision.
Friction is included by assuming that the grains have a coefficient of friction $\mu$.  If the ratio of the tangential to the normal force exceeds
$\mu$, then the grains are allowed to slip with a tangential force equal to $\mu$ times the normal force.  Using these dynamical constraints, 
contact forces can be determined at each time step. 

It is important to mention one subtlety of the algorithm: because a constant time step is utilized, many contacts can (and do) occur in each time step.  In this case, all contacts are assumed to occur simultaneously and the dynamical rule in Equation~(\ref{dynamicalrule}), along with the friction constraint, is applied to each contact.  This leads to a set of coupled algebraic equations for each group of contacting grains that is solved using an iterative method.  Therefore, the value of a specific contact force depends not only on properties of the contacting grains, but also on the properties of other grains in the connected cluster.  Physically, this corresponds to the effects of contact forces propagating through a network of grains much faster than contacts are created and/or destroyed.  The CD algorithm explores the limit of infinitely rigid grains
by assuming that the propagation occurs instantaneously.

The limit of infinitely rigid grains has also been explored in other simulations~\cite{campbellrigid,grestsilbert,dacruz, comparemdcd}.  In one study~\cite{campbellrigid}, Campbell simulated soft spheres with a rigidity $k$ in shear flow at constant volume.  He found that, for any packing fraction below a critical value $\nu_\mathrm{c}$, $k$ can be taken large enough so that no macroscopic variables depend on its value.  
This study, along with others~\cite{dacruz, grestsilbert, comparemdcd}, establishes the validity of the ``hard-sphere'' limit in granular shear flow.

Shear flows of hard-sphere granular materials always occur in the inertial regime~\cite{Lemaitreprl, gaj1}.  Because the stiffness of the grains does not play a role, the 
interactions between grains are mediated by dimensionless quantities, as in Equation~(\ref{dynamicalrule}), and the only timescale is set by
the value of the external shear rate $\dot\gamma$.  Therefore, by dimensional analysis, it must be the case that the pressure and the shear stress are proportional 
to $\dot\gamma^2$.  This is Bagnold's scaling~\cite{bagnold54}, which is a central characteristic of granular flows in the inertial regime.  Since hard-sphere granular flows must 
always occur in the inertial regime, our simulations are constrained to lie below the jamming transition at $\nu_\mathrm{c}$ and always have zero yield stress.  However, we can explore granular flows arbitrarily close to $\nu_\mathrm{c}$ and we have conducted simulations up to packing fraction of $0.84$.

The value of $\nu_\mathrm{c}$ in our system is ultimately determined by the distribution of grain sizes.  In the simulations presented here we use a polydisperse collection of circular grains in two dimensions.  The grain diameters are chosen randomly from a flat distribution with minimum and maximum diameters given by $\sigma \pm \Delta$, with $\sigma = 1.4$ and $\Delta = 0.26 \sigma$.  We have found that this amount of polydispersity restricts crystallization.  We estimate that the value of $\nu_\mathrm{c}$ for this grain polydispersity, given the shearing algorithm, is between $0.84$ and $0.85$.  

\begin{figure}
\resizebox{!}{.38\textwidth}{{\includegraphics{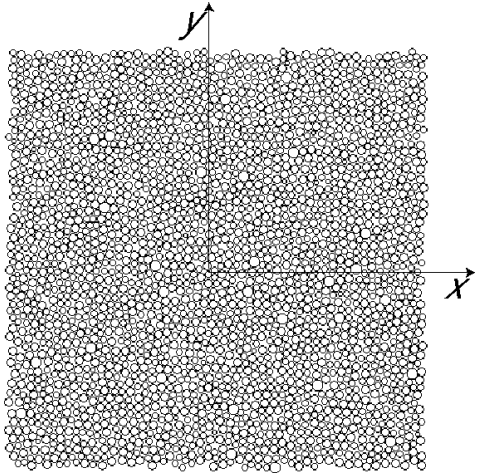}}}
\caption{\label{figsims} Snapshot of a granular simple shear flow.  Each grain has an average velocity in the x-direction given by $\dot\gamma y$, where $\dot\gamma$ is the shear rate.  The center of the cell is defined as $x=y=0$.}
\end{figure}

Finally we comment on the shearing algorithm.  We use a numerical procedure to create a simple shear flow without any boundaries.  This is done by using 
the Lees-Edwards boundary conditions~\cite{leesedwards} along with the sllod equations of motion~\cite{sllod} to initiate shear flow immediately.  We discuss
these procedures at length in an earlier paper~\cite{gaj1}.  For our purposes here, it suffices to state that this procedure produces a translationally invariant simple shear flow with a linear velocity profile.  There are no other gradients in the system and the stress tensor, granular temperature, and packing fraction are all constant.  Although this is an idealized simulation, we have shown that the resulting rheology is identical to the rheology found in steady state incline flow, far from the boundaries~\cite{gaj1}.  A screenshot of our simple shear flow simulation is shown in Figure~\ref{figsims}.    

\section{Force Correlations}
\label{xisection}

In this section we present the central measurement of the paper.  We investigate spatial force-force correlations in steady state shear flows through the measurement of the correlation function
\begin{equation}
C(\mbox{\boldmath $\ell$}) = \frac{\sum_{i=1}^N \sum_{j=1}^{i-1} {\bf F}^i \cdot {\bf F}^j \delta({\bf r}^i - {\bf r}^j - \mbox{\boldmath $\ell$})}{\sum_{i=1}^N {\bf F}^i \cdot {\bf F}^i}.  
\label{isotropicforcecorr}
\end{equation}
In this equation, ${\bf F}^i$ is the total vector force (sum of contact forces) experienced by a grain $i$ at position ${\bf r}^i$ and the sums are taken over all grains that have at least one contact.  The distance $\mbox{\boldmath $\ell$}$ ranges over the entire system size and is not limited to grains in direct contact.  We take the average value of $C(\mbox{\boldmath $\ell$})$ over at least $5000$ time steps in steady state shear flow.  A non-zero value of $C(\mbox{\boldmath $\ell$})$ reveals that, on the average, two grains separated by a distance $\mbox{\boldmath $\ell$}$ have forces that are correlated. 

For perfectly rigid granular materials, where there is only a repulsive interaction between grains {\em at contact}, $C(\mbox{\boldmath $\ell$})$ gives a quantitative measurement of the average effect of force chains of length $\mbox{\boldmath $\ell$}$ in the material.  
A non-zero value of the correlation indicates that two grains a distance $\mbox{\boldmath $\ell$}$ are connected through a cluster of simultaneously contacting grains and the force from one grain is being transmitted through the network to the other grain.  It thereby establishes that simultaneous contacts exist and that forces propagate through networks.  Positive values of the correlation correspond to situations where the total forces on each grain tend to be aligned.  

Because we make the measurement of $C(\mbox{\boldmath $\ell$})$ while the material is in steady state shear flow, the correlations do not reveal the presence of static structure.  Instead, because contacts between grains are being created and destroyed by the overall flow, the correlation function gives information on the average size of dynamic structures that are fluctuating in both space and time.

We will demonstrate that the correlation function depends on the vector distance $\mbox{\boldmath $\ell$} = \ell \mbox{\boldmath $\hat{\ell}$}$ between pairs of grains, and that it decays exponentially with $\ell$.  In the following sections, we first investigate the dependence of $C(\mbox{\boldmath $\ell$})$ on the magnitude of $\mbox{\boldmath $\ell$}$, thereby defining an isotropic correlation length $\xi$.  Then we investigate the full dependence of $C(\mbox{\boldmath $\ell$})$ and obtain the full functional form of $\xi(\theta)$, which depends on the angle $\theta$ between pairs of grains. 

\subsection{Measurements of $C(\ell)$ and the length scale $\xi$}
\begin{figure}[htbp]
\begin{center}
\psfrag{yl}{\Huge{$C(\ell)$}}
\psfrag{xl}{\Huge{$\ell/\sigma$}}
\subfigure{\scalebox{0.37}{\includegraphics{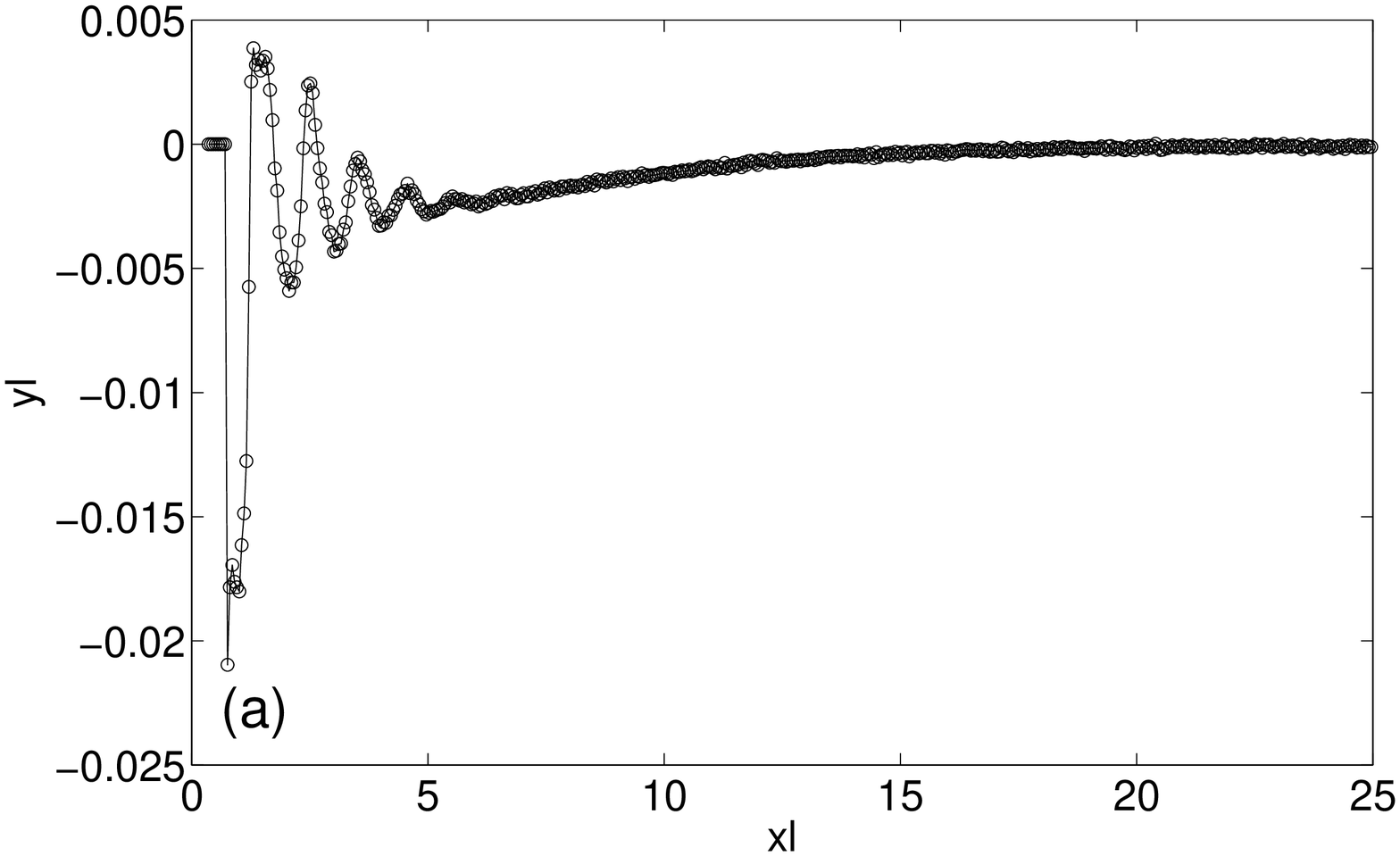}}}
\quad
\psfrag{yl}{\Huge{$\log |C(\ell)|$}}
\psfrag{xl}{\Huge{$\ell/\sigma$}}
\subfigure{\scalebox{0.37}{\includegraphics{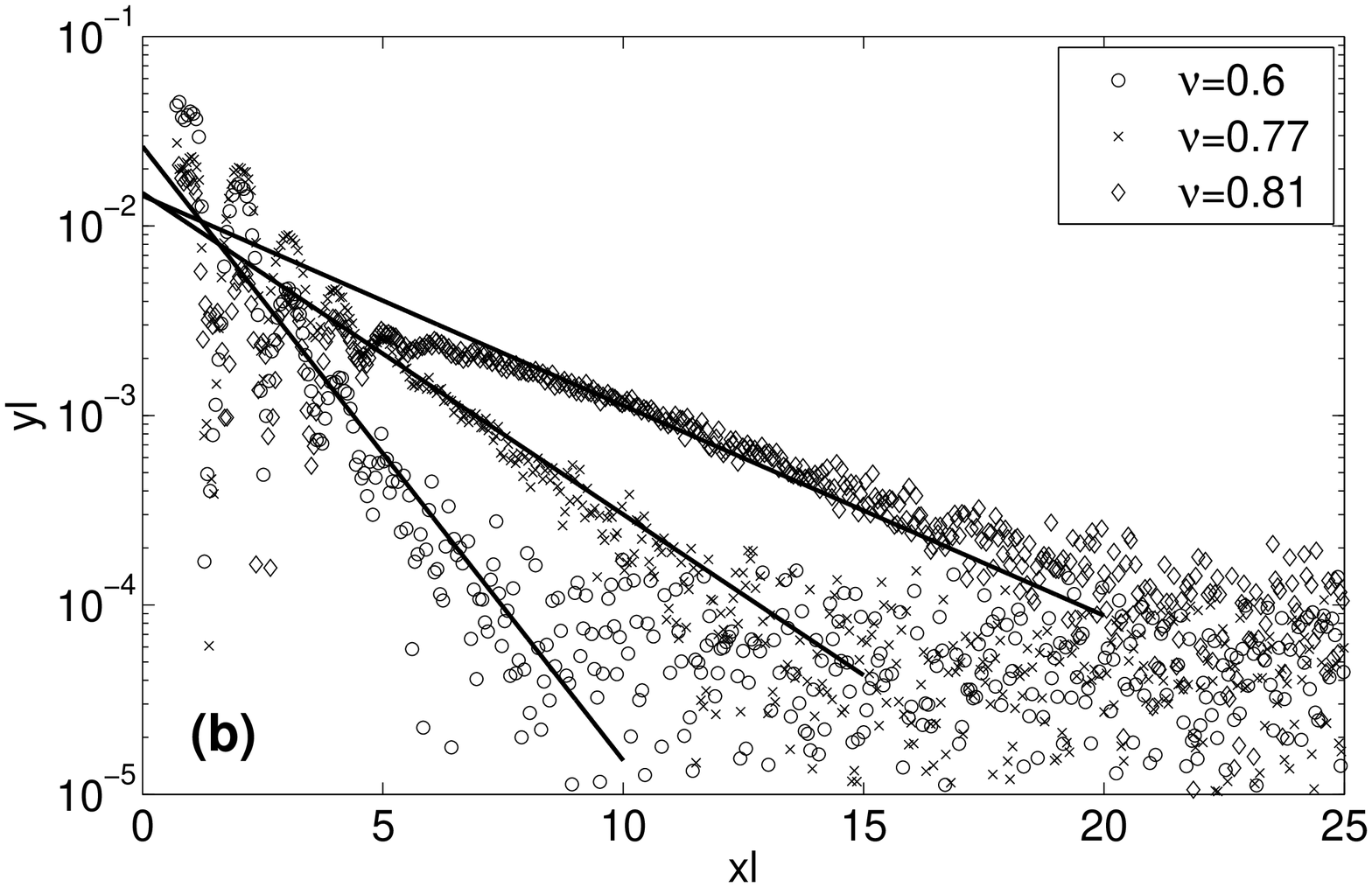}}}
\caption{\label{figcl} {\bf (a)} The force correlation function $C(\ell)$ for $e=0.25$
and $\nu = 0.81$, where $\ell$ is the distance between grains and $\sigma$ is the average
grain diameter.  {\bf (b)}  The logarithm of the magnitude, $\log |C(\ell)|$, for packing fractions
of $\nu=0.6$, $0.77$, and $0.81$,
illustrating the exponential decay of the correlations.  The lines correspond to the
function $e^{-\ell/\xi}$, where $\xi$ is determined from Equation~(\ref{eqnforxi}) and plotted in Figure~\ref{figxi}.
}
\end{center}
\end{figure}

We begin by measuring the isotropic part of the correlation $C(\ell)$ in two dimensional simulations by averaging $C(\mbox{\boldmath $\ell$})$ over all directions \mbox{\boldmath $\hat{\ell}$}.
In Figure~\ref{figcl} we plot this directionally averaged correlation function $C(\ell)$ for a frictionless material with $e=0.25$.  

The logarithm of the magnitude of the correlation function $\log |C(\ell)|$ is also plotted in Figure~\ref{figcl}.  From these plots we observe that the magnitude decreases exponentially, although the decay is complicated by an oscillating function that accounts for the sign of $C(\ell)$.  The form of this oscillating function is not universal and depends on the exact value of density and restitution.  Nevertheless, as a first approximation, we express the correlation function as $C(\ell) \sim \exp(-\ell/\xi)$, which introduces a length scale $\xi$ that quantifies the large $\ell$ behavior of the correlations.  The data in Figure~\ref{figcl} illustrates that this length scale increases with packing fraction.             

We find that the value of $\xi$ is well approximated by the equation
\begin{equation}
\xi = \frac{\int_0^{\infty}{ d\ell \, \ell \, C(\ell)}}{\int_0^{\infty}{ d\ell \, C(\ell)}}.
\label{eqnforxi}
\end{equation}
In Figure~\ref{figcl} we plot $\exp(-\ell/\xi)$, where $\xi$ is determined for each density from 
Equation~(\ref{eqnforxi}), and we observe excellent agreement with the measured exponential decays of $C(\ell)$.
Although Equation~(\ref{eqnforxi}) has the same form as an expectation value in statistical physics, that is not the proper interpretation here.  Rather, Equation~(\ref{eqnforxi}) is simply a combination of integrals that gives the coefficient of any exponential function.
  
\begin{figure}[htbp]
\begin{center}
\psfrag{yl}{\Huge{$\xi/\sigma$}}
\psfrag{iyl}{\Huge{$\xi/\sigma$}}
\psfrag{xil}{\Huge{$\nu$}}
\subfigure{\scalebox{0.37}{\includegraphics{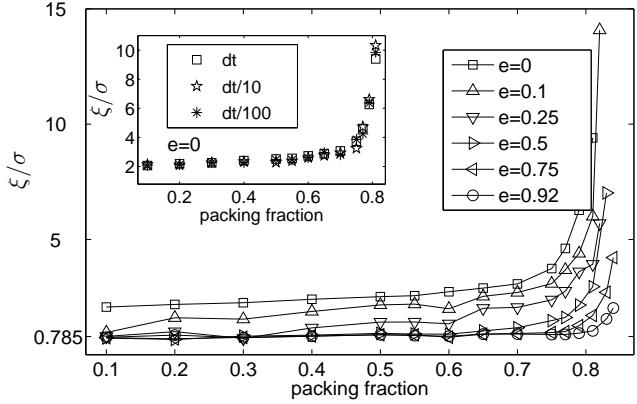}}}
\caption{\label{figxi} {\bf Main Figure:} The length scale $\xi$ for frictionless granular shear flows, normalized by the average radius $\sigma$, plotted for a wide range of packing fraction and restitution $e$.  The value of $\xi$ increases rapidly for large packing fraction and asymptotes to $\xi_\mathrm{el}/\sigma \equiv 0.785$ for small packing fraction.  {\bf Inset:} Data for $e=0$ at multiple values of the time step.  Here $dt$ corresponds to the time step we use in this paper, which is smaller for larger packing fractions.  The value of $\xi$ does not depend on the time step.  }
\end{center}
\end{figure}

In Figure~\ref{figxi} we plot measurements of $\xi$, determined from Equation~(\ref{eqnforxi}), for all of the frictionless granular flows we have simulated.  The value of $\xi$ quantifies the average extent of force chains in the system and 
is an increasing function of density for each value of restitution.  
Higher values of restitution have lower values of $\xi$.  This is consistent with dynamic force networks forming spontaneously as the material is sheared due to energy dissipation upon contact.  Indeed, in the limit of $e \rightarrow 1$, the data in Figure~\ref{figxi} suggests that no correlations exist.  The formation of correlations is therefore related to the observation of inelastic collapse in sheared granular materials~\cite{hrenyacollapse}.  Collapsing grains contact increasingly often, allowing a single grain to have multiple contacts, even in the limit of perfectly rigid grains~\cite{gaj2}.    

For large values of packing fraction, the data for $\xi$ increases rapidly.  The maximum packing fraction we are able to simulate is $\nu_{\mathrm{max}} = 0.84$, but we were not able to determine the value of $\xi$ at $\nu_{\mathrm{max}}$ for all values of restitution.  This is because the length scale of the correlations becomes larger than the maximum system size of $100 \times 100$ grains that we are able to simulate.  For the high values of packing fraction, we conduct simulations for packing fractions in increments of $0.01$.  When $e \leq 0.5$ we reach a packing fraction where, if we increase the packing fraction by 0.01, $\xi$ is too large to be measured.  The rapid growth of the correlations at high packing fraction suggests that $\xi$ is diverging.

For small values of packing fraction, Figure~\ref{figxi} shows the value of $\xi$ approaching a limiting value of $0.785 \sigma$.  This corresponds to the smallest correlation possible, which arises when the only relevant interactions are binary collisions between grains.  For these small packing fractions, the sign of $C(\ell)$ is exclusively negative and the exponential decay is not observed.  Indeed, in the limited simulations we have conducted with a mono-disperse collection at very low packing fraction, we observe that $C(\ell)$ is zero for all values of $\ell$ except $\ell = \sigma$.  
In the case of a polydisperse collection of grains, the form of $C(\ell)$ for dilute flows depends on the relative probabilities to have binary interactions between grains of different sizes and exponential decay is not observed since the correlation function equals zero for all values of $\ell$ larger than the maximum grain diameter.  Even though exponential decay is not observed in the limiting case of binary collisions between grains, Equation~(\ref{eqnforxi}) still provides a useful measure of correlation length that matches the exponential decay at higher densities.

\begin{figure}[htbp]
\begin{center}
\psfrag{yl}{\Huge{$\xi_{\mathrm{el}}/\sigma$}}
\psfrag{xl}{\Huge{$\Delta/\sigma$}}
\subfigure{\scalebox{0.37}{\includegraphics{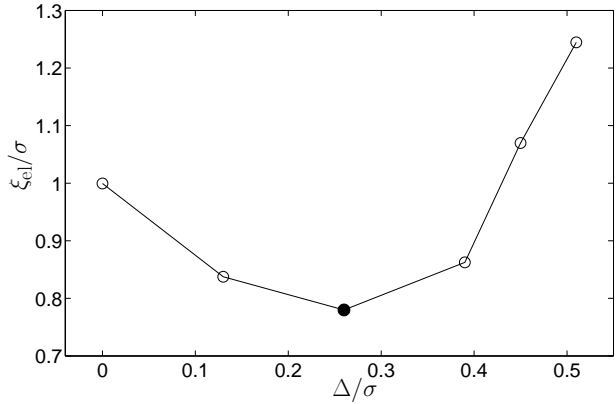}}}
\caption{\label{xielastic} The equilibrium correlation length $\xi_{\mathrm{el}}$ as a function of the grain distribution of the system.  We consider flat distributions with minimum and maximum grain diameter given by $\sigma \pm \Delta$.  The distribution considered in this paper is indicated by a solid circle.}
\end{center}
\end{figure}

The limiting value of $0.785$ is related to the probability to have a binary collision between grains of different diameters and is equal to the correlation length that would be expected in a perfectly elastic system where $e=1$ and no energy is dissipated upon interactions between grains.  Thus we denote the limiting value as $\xi_{\mathrm{el}}$ and expect that its numerical value is related to the distribution of grain sizes.  We have conducted a limited number of simulations to determine the behavior of $\xi_{\mathrm{el}}$ for different grain distributions, and the results are plotted in Figure~\ref{xielastic}.  Each grain distribution we consider is a flat distribution with minimum and maximum grain diameter given by $\sigma \pm \Delta$.  As expected, $\xi_{\mathrm{el}}/\sigma=1$ for $\Delta=0$.  For large values of $\Delta$, $\xi_{\mathrm{el}}$ depends linearly on $\Delta$ since the largest grains set the correlations.  The grain distribution we consider in this paper corresponds to $\Delta = 0.26 \sigma$ and is designated by a solid data point.  Although this grain distribution happens to occur close to the minimum, this is not relevant to the arguments in this paper.  We present the data in Figure~\ref{xielastic} to better understand the origin of $\xi_\mathrm{el}$, and in particular why it is not always equal to one.  We will see that the ratio $\xi/\xi_\mathrm{el}$ seems to signal the transition between different regimes of shear flow, and it asymptotes to one for small packing by definition.

\begin{figure}[htbp]
\begin{center}
\psfrag{yl}{\Huge{$\xi/\sigma$}}
\subfigure{\scalebox{0.37}{\includegraphics{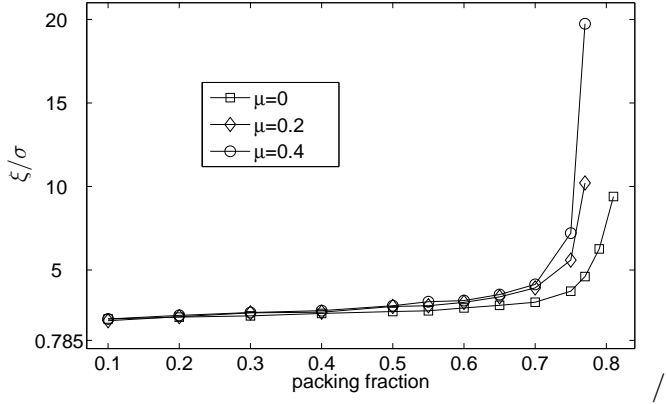}}}
/\caption{\label{figmuxi} The length scale $\xi$, normalized by average diameter $\sigma$, for granular shear flows with friction coefficient $\mu$ and $e=0$.  Larger values of $\mu$ correspond to larger $\xi$.  The value of $\xi/\sigma$ is always larger than $\xi_\mathrm{el}/\sigma=0.785$}
\end{center}
\end{figure}

The data we have presented for frictionless grains can also be extended to situations where $\mu>0$.
Adding friction introduces a non-zero tangential force at each contact that tends to increase geometrical frustration.
Measurements of $\xi$ for frictional systems are plotted in Figure~\ref{figmuxi} for $e=0$ and three friction coefficients $\mu$.  We see that the value of $\xi$ diverges sooner for systems with friction.  This is to be expected, since the additional constraints associated with the tangential forces can only increase the correlations.  Interestingly, the value of $\xi$ for different friction coefficients is not different below a rather large packing fraction.  This is because clusters of simultaneously contacting grains must exist before the sticking effects of friction alter the dynamics and correlations.  
 From a qualitative viewpoint, the behavior of $\xi$ for frictional systems is analogous to the frictionless case-- the length-scale is observed to diverge at finite packing fraction and asymptote to $\xi_\mathrm{el}$ at low packing.

\subsection{Connection to Jamming}
The data in Figures~\ref{figxi}~and~\ref{figmuxi} suggests that $\xi$ diverges at a finite packing fraction $\nu_\mathrm{c}$ that may depend on the friction coefficient and restitution coefficient.
This divergence is related to the jamming transition~\cite{liunagel} in granular materials, where the shear modulus becomes non-zero and the system is able to sustain a shear stress without yielding.
In order to make the transition from a flowing shear state to a jammed state, it is necessary that a correlation length approach the size of the system, and diverge in the thermodynamic limit.  This is because force chains must percolate from the upper to lower shearing wall in order to counteract the applied shearing force.  The correlation length $\xi$ quantifies the notion of force chains and we expect that the observed divergence of $\xi$ is a necessary condition for jamming. 

However, there is no guarantee that the divergence of $\xi$ is also a sufficient condition for jamming, since it is possible that force chains percolate long before the system jams.  Nevertheless, both theories~\cite{moukarzel1,moukarzel2} and simulations~\cite{aharanovsparks, ohernsilbert, zhangmakse} have found that percolation and jamming occur simultaneously, which suggests that a granular system jams if, and only if, $\xi$ diverges. 
If this holds, then we would expect the jamming transition to occur at lower packing fraction as the friction between grains increases.


\subsection{Anisotropy in the Angular Dependence of $\xi$}
\begin{figure}[htbp]
\begin{center}
\subfigure{\scalebox{0.37}{\includegraphics{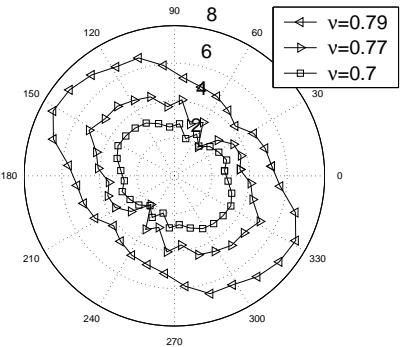}}}
\caption{\label{e0anglexi} The angular dependence of the length scale $\xi(\theta)$ for a frictionless shear flow with $e=0$ and three different packing fractions $\nu$.  The length scale is greatest along the compressional axis of the shear flow.
}
\end{center}
\end{figure}

We have also conducted measurements of the full directional dependence of $C(\mbox{\boldmath $\ell$})$.  In this case, we observe that the decay of the correlations can still be described by an exponential, but the value of the correlation length depends on the orientation \mbox{\boldmath $\hat{\ell}$}.  In two dimensions this orientation can be quantified by the angle $\theta$ between \mbox{\boldmath $\ell$} and the $x$-axis.  In Figure~\ref{e0anglexi} we plot the angular dependence of $\xi$ for frictionless flows with $e=0$ and three different packing fractions.  These plots reveal that $\xi$ is not isotropic and the correlation between a pair of grains depends on their orientation.  For a simple shear flow, correlations are created along the compressional axis due to a higher number of grain interactions, and is destroyed along the dilational axis due to contacts being lost.      

\begin{figure}[htbp]
\begin{center}
\subfigure{\scalebox{0.37}{\includegraphics{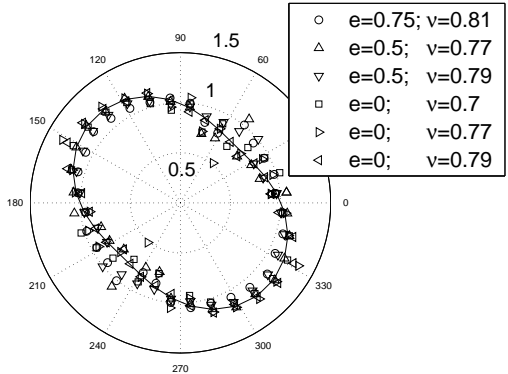}}}
\caption{\label{anglexicollapse} The normalized angular dependence of the length scale, $\xi(\theta)/\xi$ for frictionless shear flows with different restitution coefficients and packing fractions.  The data is well characterized by Equation~(\ref{collapseeqn}), plotted as a line.
}
\end{center}
\end{figure}

We note that the maximum value of the correlation length occurs at approximately the same angle for each packing fraction in Figure~\ref{e0anglexi}.  This trend is followed for other packing fractions and restitution coefficients as well.  In Figure~\ref{anglexicollapse} we plot the angular dependence of the length scale divided by its average value: $\xi(\theta)/\xi$.  We notice that this collapses the data for a large range of packing fractions and restitution coefficients onto one curve.  The collapse is not perfect, especially along the dilational axis of the shear flow where the correlations are small.  We believe this to be due to the small number of collisions that occur on this axis, which makes gathering statistics difficult.  

The common collapsed curve for all of the data in Figure~\ref{anglexicollapse} shows that $\xi$ is anisotropic.
The angular dependence of $\xi$ can be approximated as a Fourier series, only including terms that are $\pi$-periodic.  We find that, as is the case for other anisotropies in granular flow~\cite{thornton,arthur,rothenburg,anisradjai,kruyt, radjai3}, the functional form of $\xi(\theta)$ is well characterized by the first two terms in the Fourier series
\begin{equation}
\xi(\theta) = \frac{\xi}{2 \pi} \left(1-a_0 \sin{2[\theta-\theta_0]} \right),
\label{collapseeqn}
\end{equation}
where $\xi$ is the average value of the correlation length.  The solid curve in Figure~\ref{anglexicollapse} is a fit to this equation, which estimates the parameter values as $a_0 = 0.21$ and $\theta_0 = 0.013$.  The value of $\theta_0$ is consistent with the axis of maximum compression for the data we have gathered.  This implies that larger values of $\xi$ occur near $\theta = 3 \pi /4$ because the compression induced by the shear flow causes more grains to come into contact.  Near $\theta = \pi/4$, where $\xi$ is a minimum, dilation reduces the magnitude of correlations and thereby the length scale.  However, we have found no simple explanation for the value of $a_0$.  We suspect that the numerical value of $a_0$ should be related to the anisotropy in the contact distribution~\cite{anisradjai}.  However, because the anisotropy in $\xi$ remains constant over a range of packing fractions and restitution coefficients where the anisotropy in the contact distribution is not constant, the value of $a_0$ may have a simpler origin.

\section{The Effects of Correlations on Contact Forces}
\label{forcepdfsection}
In the previous section we presented measurements of a correlation length $\xi$ that diverges at the jamming transition and asymptotes to an elastic value $\xi_\mathrm{el}$ at small packing fraction.  This length scale captures the decay of force correlations and is related to the emergence of clusters of simultaneously contacting grains.  In very dilute systems only binary collisions are relevant, $\xi = \xi_\mathrm{el}$, and contact forces can be determined from the properties of the two colliding grains.  However, as the packing fraction is increased, $\xi$ also increases and the contact force between any two grains will depend on properties of the other grains in the cluster.  This is because, in the rigid grain limit, forces propagate instantaneously through the network.  Because the growth of $\xi$ is closely related to the nature of the force transfer-- binary collisions or force networks-- we expect the contact forces to depend on the value of $\xi$.  A useful way to explore properties of contact forces is to measure the contact force probability distribution function (pdf) $P(F)$.  This function encodes the statistics of the contact forces: $P(F) dF$ is proportional to the number of contact forces in the range $F$ to $F+dF$.  

To make the connection between contact forces and long range spatial force correlations explicit, 
we begin by demonstrating that the contact forces between pairs of grains can not be determined simply by assuming binary collisions when $\xi$ is large.  
To illustrate this point, we compare the statistics of the actual contact forces to the forces we would calculate if we assumed that only binary collisions occurred.  If we make the binary collision assumption, then the dynamical rule in Equation~(\ref{dynamicalrule}), along with momentum conservation, determines the normal impulse in each collision.  Dividing this impulse by the algorithm time step yields the average force that would arise over the time step $dt$.  We label this force $F_\mathrm{bc}^{ij}$, where $i$ and $j$ represent the colliding grains and the label ``$\mathrm{bc}$'' reminds us that this force only applies to purely binary collisions.  It is simple to show that the value of the binary force is given by 
\begin{equation}
F^{ij}_{\mathrm{bc}} = (1+e) \mu^{ij} \left[ ({\bf v'}^j - {\bf v'}^i) \cdot \mbox{\boldmath $\hat{\sigma}$}^{ij} \right]/dt,
\label{binaryforce}
\end{equation}
where $e$ is the normal restitution coefficient, $\mu = m^i m^j/(m^i+m^j)$ is the reduced mass, ${\bf v'}^i$ the pre-collisional velocity of grain $i$, and $\mbox{\boldmath $\hat{\sigma}$}^{ij}$ is the unit vector connecting the centers of grains $i$ and $j$.  All of these terms can be
measured directly from simulations. 

\begin{figure}[htbp]
\begin{center}
\mbox{
\psfrag{yl}{\Huge{$\log P(f)$}}
\psfrag{xl}{\Huge{$f$}}
\scalebox{0.37}{\includegraphics{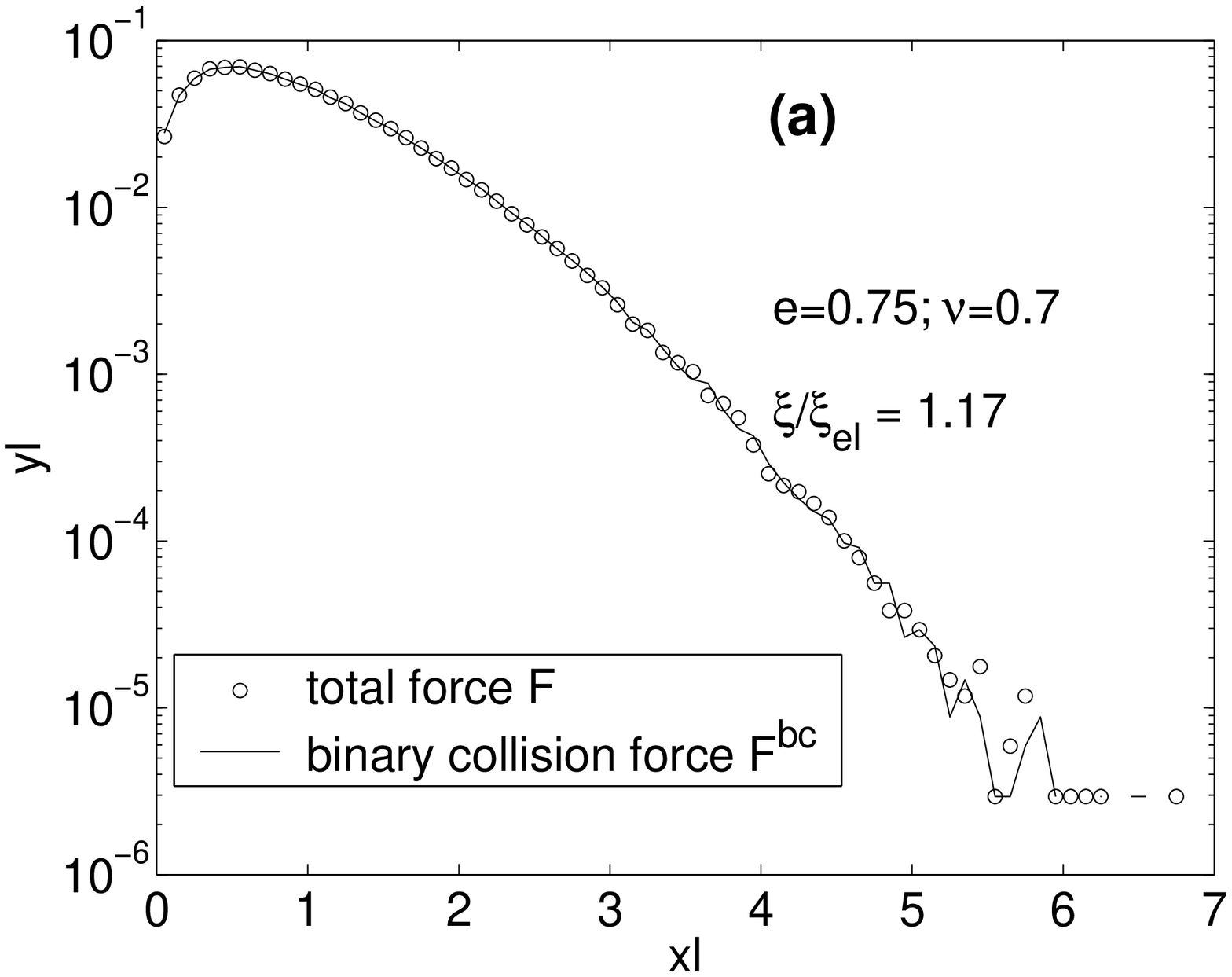}}
}
\mbox{
\psfrag{yl}{\Huge{$\log P(f)$}}
\psfrag{xl}{\Huge{$f$}}
\scalebox{0.37}{\includegraphics{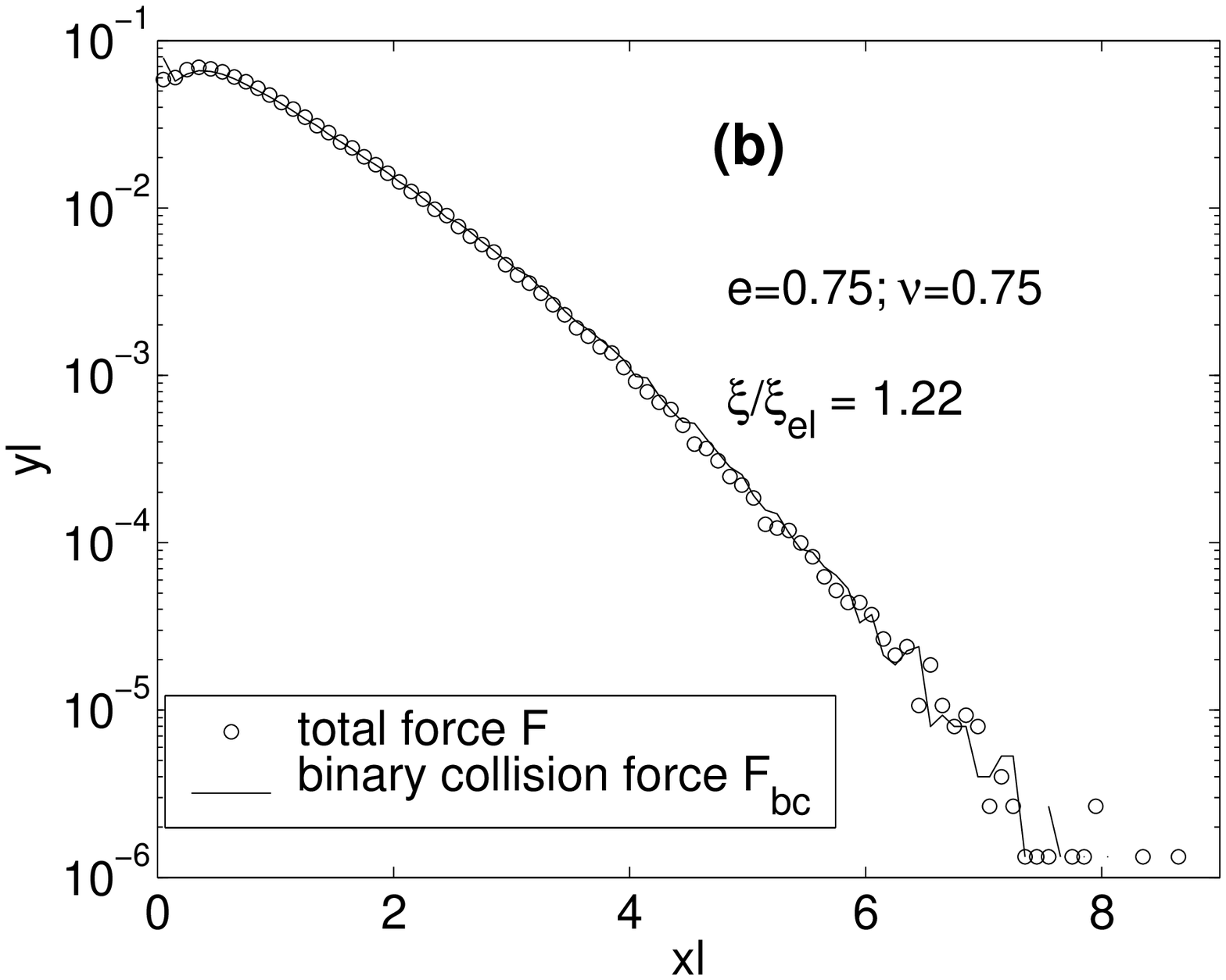}}
}
\mbox{
\psfrag{yl}{\Huge{$\log P(f)$}}
\psfrag{xl}{\Huge{$f$}}
\scalebox{0.37}{\includegraphics{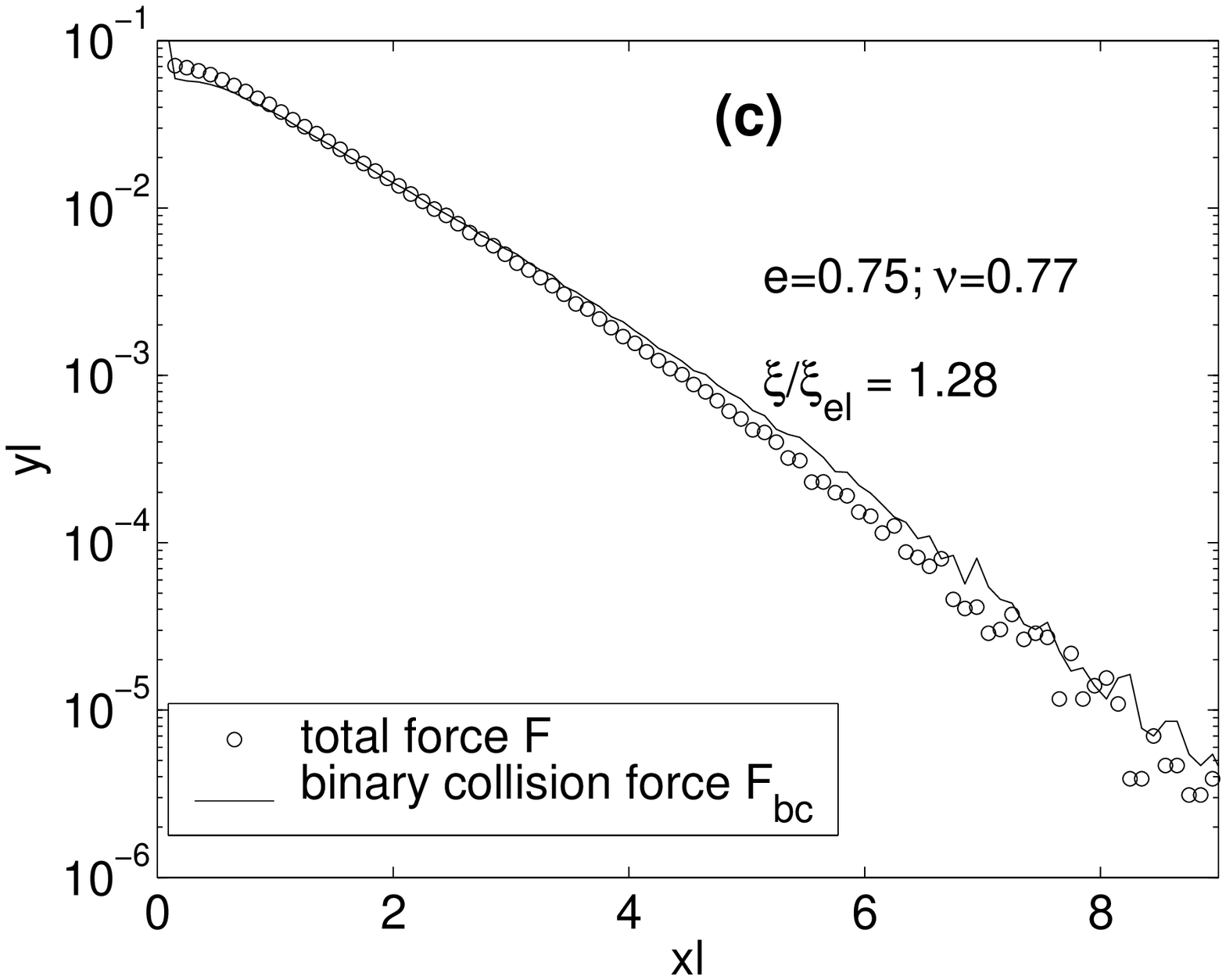}}
}
\mbox{
\psfrag{yl}{\Huge{$\log P(f)$}}
\psfrag{xl}{\Huge{$f$}}
\scalebox{0.37}{\includegraphics{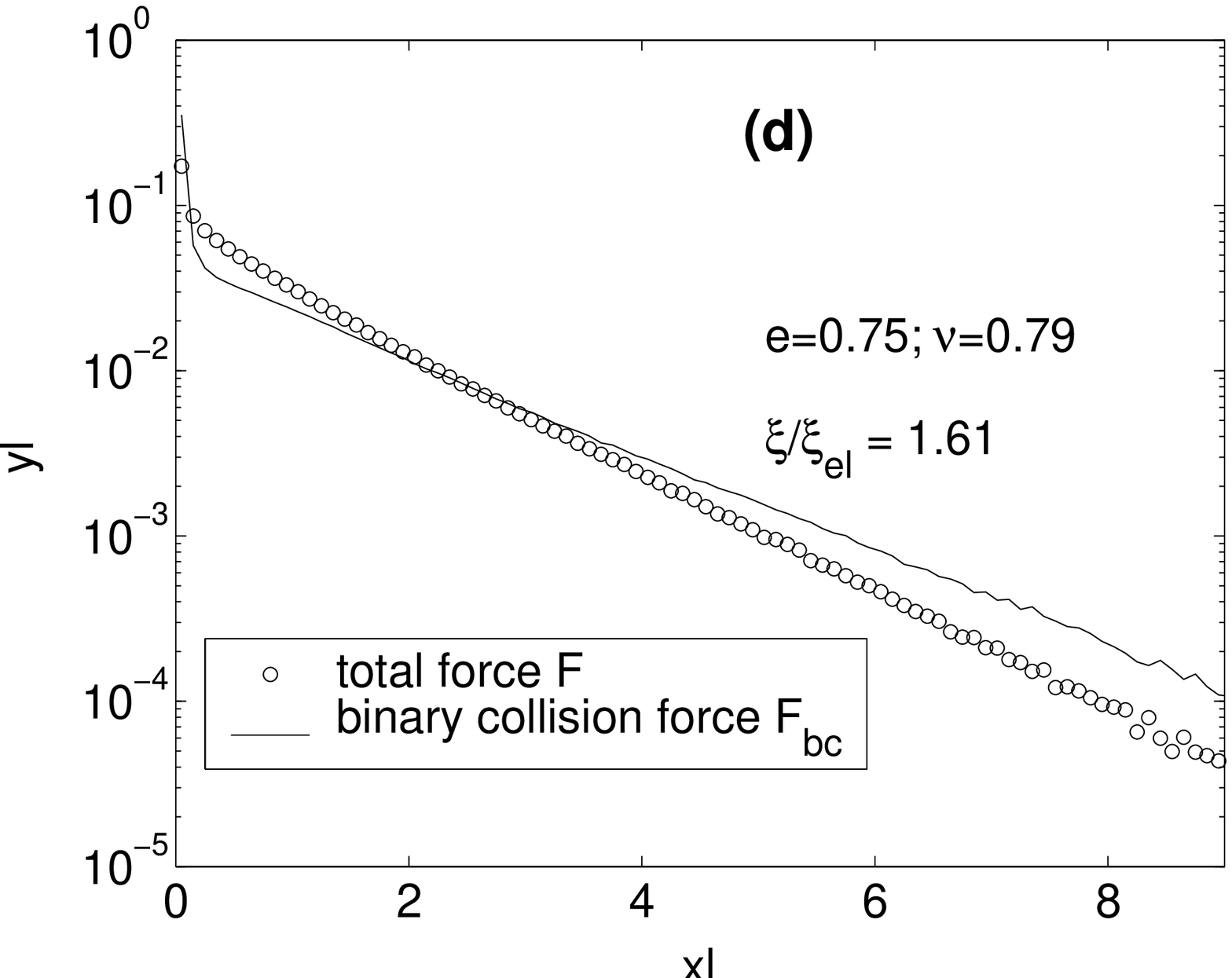}}
}
\mbox{
\psfrag{yl}{\Huge{$\log P(f)$}}
\psfrag{xl}{\Huge{$f$}}
\scalebox{0.37}{\includegraphics{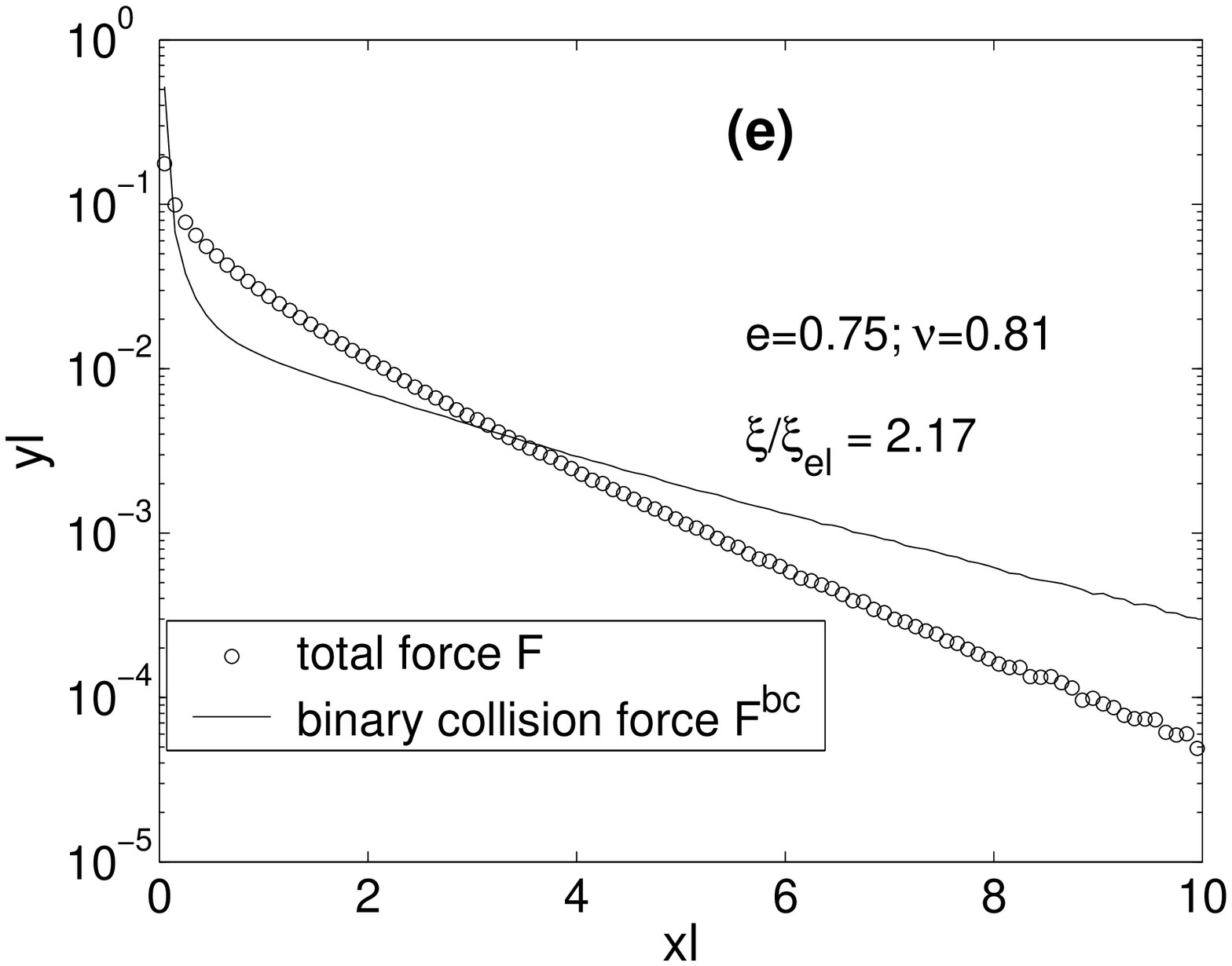}}
}
\mbox{
\psfrag{yl}{\Huge{$\log P(f)$}}
\psfrag{xl}{\Huge{$f$}}
\scalebox{0.37}{\includegraphics{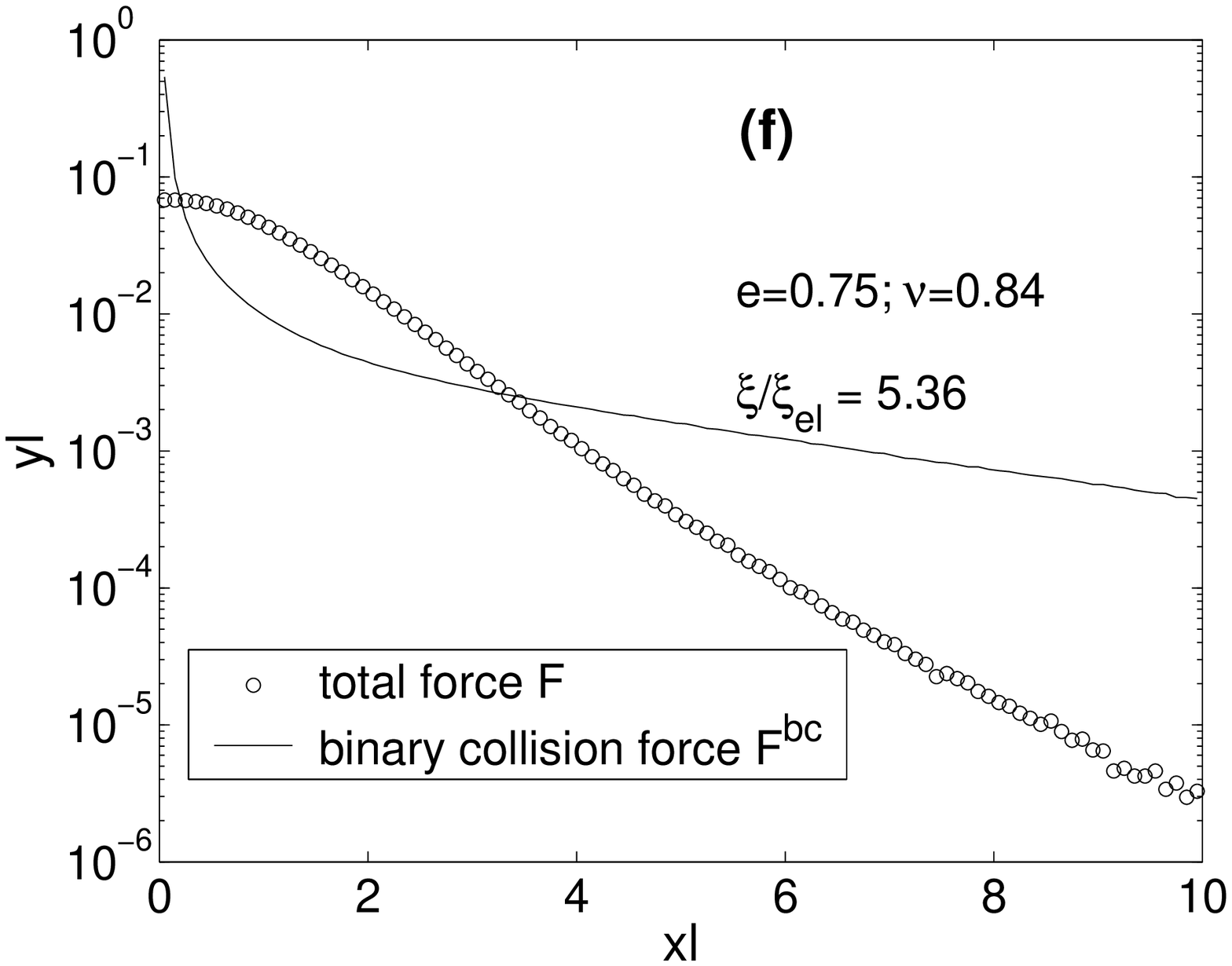}}
}
\caption{\label{figkineticpdf} The data points are $P(f)$ for systems with $e=0.75$ and growing values of $\xi/\xi_{\mathrm{el}}$, where $f$ is the contact force $F$ divided by the average value $\langle F \rangle$.  This data is compared with the line, where the force is determined from Equation~(\ref{binaryforce}) which assumes that only binary collisions occur.  We have normalized both the total forces and binary collision forces in each plot by their average values.  There is excellent agreement for $\xi/\xi_\mathrm{el} < 1.25$.  For larger values of the correlation length, clusters of interacting grains form and the binary collision assumption does not fit the data.
}
\end{center}
\end{figure}

In Figure~\ref{figkineticpdf} we plot measurements of the contact force distribution function $P(F)$ for six different values of restitution and packing fraction.  In each figure we compare $P(F)$ with the statistics of the binary forces $P(F_\mathrm{bc})$.  If these two functions are the same then contact forces are well approximated by only considering binary collisions; if the functions differ then we know that clusters of contacting grains affect contact forces.

We indicate the value of $\xi/\xi_\mathrm{el}$ for each plot in Figure~\ref{figkineticpdf} and immediately see that for 
small values of $\xi/\xi_\mathrm{el}$ the data for $P(F)$ is well fit by the line, which is a measurement of $P(F_\mathrm{bc})$.  However, as $\xi$ increases, the presence of force networks changes the nature of the contact forces and we can no longer make accurate predictions by assuming that only binary collisions occur.  The value $\xi/\xi_\mathrm{el} \approx 1.25$ serves as a rough upper bound for the regime where the binary collision assumption is reasonable.  Similar values of $\xi/\xi_\mathrm{el}$ signal the transition between collisional and non-collisional flows for all restitution coefficients and packing fractions we have investigated.  

This is not surprising since $\xi/\xi_\mathrm{el} > 1.25$ comprises a region where force networks have formed and simultaneous contacts occur.  In this regime, in order to calculate the force between two grains, it is not sufficient to only consider the properties of the two contacting grains.  Rather, all of the grains connected in the force network play an important role.  This is because the two contacting grains are being pushed together by the other grains in the cluster and the contact force is equal to the binary collision contribution from Equation~(\ref{binaryforce}) {\em plus} a contribution from the cluster.  

We conclude from Figure~\ref{figkineticpdf} that $\xi/\xi_\mathrm{el} \approx 1.25$ separates the region where only binary collisions occur from the region where force networks begin to form and affect contact forces.  This is an approximate value of the length scale at which there is a clear deviation between binary collisional forces and total forces.  Measurements presented later in this paper, and in the companion paper~\cite{gajcompanion2}, also place the network transition at $\xi/\xi_\mathrm{el} \approx 1.25$.  Although this value of the length scale does not change with restitution coefficient or friction, we would expect it to depend on the grain size distribution.    

The techniques we used to determine the crossover in Figure~\ref{figkineticpdf} can only be used in simulations where the position, velocity, and force on every grain is always known.  However, we have also found a signature of the transition that can (and has been) observed in experiments of granular flows.  
This signature relates to the small force behavior of the contact force distribution function. 

\begin{figure}[htbp]
\begin{center}
\mbox{
\psfrag{yl}{\Huge{$\log P(f)$}}
\psfrag{xl}{\Huge{$f$}}
\scalebox{0.37}{\includegraphics{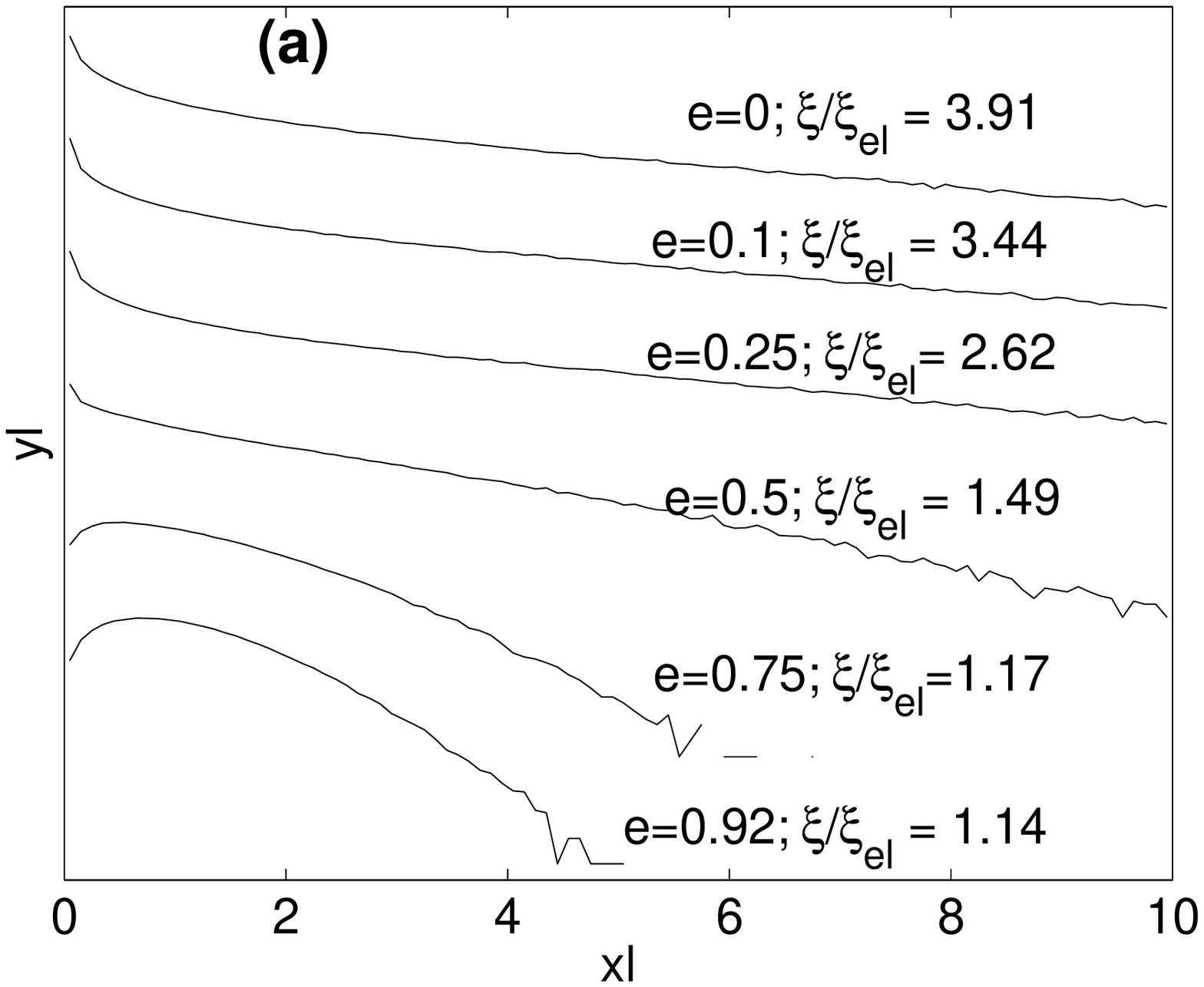}}
}
\mbox{
\psfrag{yl}{\Huge{$\log P(f)$}}
\psfrag{xl}{\Huge{$f$}}
\scalebox{0.37}{\includegraphics{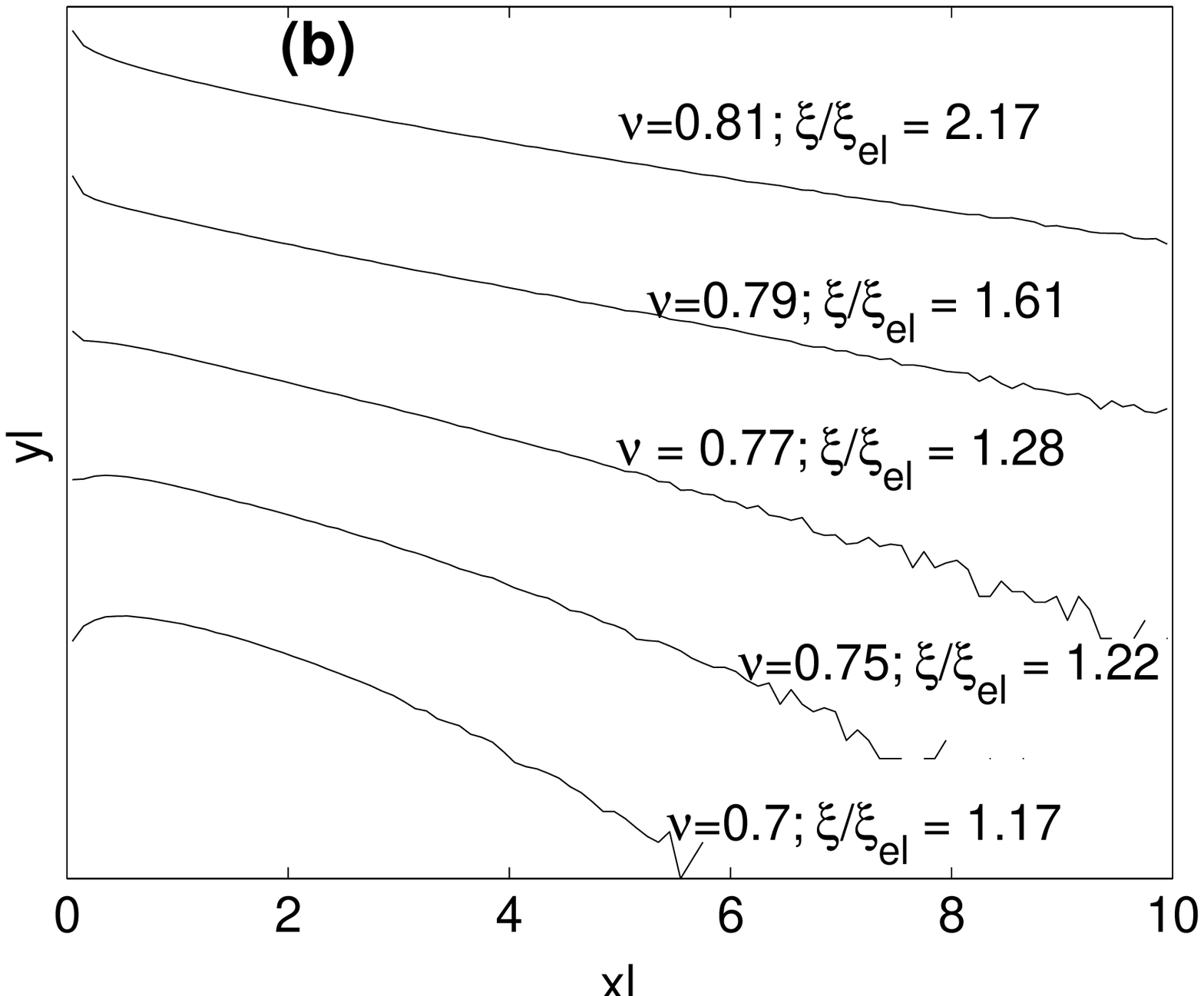}}
}
\caption{\label{forcedists} Measurements of the contact force probability distribution function $P(f)$, where $f$ is the value of the contact force divided
by the average contact force.  The different curves have been offset so each curve is easy to see.  {\bf (a)} Data for $\nu=0.7$ and different values of the restitution coefficient $e$.  {\bf (b)} Data for $e=0.75$ and increasing values of packing fraction $\nu$.  The curves in each plot are also labeled by their associated value of $\xi/\xi_{\mathrm{el}}$, and we observe that the peak is present in $P(f)$ only if $\xi/\xi_{\mathrm{el}}<1.28$.
}
\end{center}
\end{figure}

In Figure~\ref{forcedists} we present data for the contact force distribution function $P(F)$.
In particular, we plot $\log P(f)$ for many different values of the restitution coefficient and packing fraction, where $f$ is equal to the normal contact force $F$ divided by the average normal force $\langle F \rangle$.  All of these curves correspond to frictionless materials, but we have observed that the statistics of the normal forces display the same behavior for frictional systems.  Our measurements of $P(f)$ have been averaged over $5000$ time steps in steady state shear flow, and the different curves are vertically displaced in the figure so each can be clearly seen.  Each curve is labeled by the value of $\xi/\xi_{\mathrm{el}}$ and we immediately see that the behavior at small $f$ depends on the value of $\xi$.  For $\xi/\xi_{\mathrm{el}} \leq 1.25$ there is a clear peak, whereas for $\xi/\xi_{\mathrm{el}}\geq 1.25$, the peak disappears and the maximum occurs at $f=0$.  

This measurement once again defines a crossover around $\xi/\xi_\mathrm{el} = 1.25$.  This is the transition where the microscopic interactions change from being dominated by binary collisions between grains to being dominated by clusters of grains, forming force networks of size $\xi/\xi_\mathrm{el}$.  
When this network transition occurs, the peak disappears and the most likely force is no longer equal to the average force.  This is because grains have spontaneously formed into transient clusters and the greatest number of contacts are simply rolling over each other, which produces a very small normal force.  This moves the peak to $f=0$ once the transition has fully developed and the average force is not representative of most of the forces.  Additionally, as force-networks become long-ranged, the data shows that there is a greater probability of large forces, which arise from a large number of grains in a cluster compressing two contacting grains.

\begin{figure}[htbp]
\begin{center}
\psfrag{yl}{\Huge{$f_\mathrm{max}$}}
\psfrag{xl}{\Huge{$f$}}
\scalebox{0.37}{\includegraphics{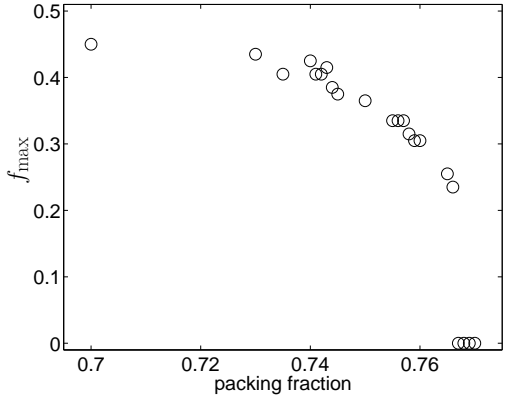}}
\caption{\label{fmaxclose} 
The value $f_\mathrm{max}$ at which the contact force pdf $P(f_\mathrm{max})$ is a maximum, as a function of packing fraction, for $e=0.75$.  The maximum of $P(f)$ occurs at positive $f$ for $\nu \leq 0.766$ and at $f=0$ for $\nu \geq 0.767$.
}
\end{center}
\end{figure}

We have also taken data of $P(F)$ close to the network transition point for granular flows with $e=0.75$.  In Figure~\ref{fmaxclose} we plot the value of the force $F_\mathrm{max}$ for which $P(F_\mathrm{max})$ is a maximum, as a function of packing fraction.  Between packing fractions of $0.766$ and $0.767$, our data shows that the value of $f_\mathrm{max}$ (equal to $F_\mathrm{max}$ divided by the average force $\langle F \rangle$) makes a jump from approximately $0.2$ to zero.  Within this small range of packing fraction, the contact force pdf loses its peak.  The abruptness of the transition suggests that different physical processes are occurring on either side of the transition and that the network transition may be quite sharp.

The signature of the transition evident in our measurements of $P(f)$ has also been observed in other simulations and experiments, but has never been connected to the formation of large scale structure.  Simulations of granular hopper flow~\cite{denniston, bulbul}, conducted using Event Driven simulations where only binary collisions are allowed to occur, have reported that as the hopper aperture is reduced and the density of the packing increases, $P(f)$ begins to lose its peak.  This is consistent with our results from Figure~\ref{forcedists} and suggests that the correlation length $\xi$ is relevant in more than just shear flows of granular media.  Additionally, the same behavior in $P(f)$ has been observed in experiments on hopper flow~\cite{hopperexps}, which lends credibility to the result and suggests that it is not an artifact of the simulation methods used here and in Refs.~\cite{denniston, bulbul}, but rather a real effect in granular flows. 

The results we have cited from hopper flow, and others~\cite{landrygrest}, have been used to challenge the belief that the formation of a peak in $P(f)$ is a signature of the jamming transition~\cite{silbertpdf, ohernpdf}.  In a wide variety of contexts, including incline flow~\cite{silbertpdf}, quasi-static flow~\cite{radjai1,radjai2}, and jammed granular materials~\cite{mueth98, lovoll, silbertgrest, landrygrestsilbert} it has been observed that $P(f)$ exhibits a maximum at $f=0$ (no peak) if the system is flowing, while a peak at non-zero $f$ forms as the systems jams.  These observations, coupled with similar results in Lennard-Jones glasses and foams~\cite{ohernpdf}, have been used to bolster the claim that the formation of a peak in $P(f)$ is a generic characteristic of the jamming transition, and a necessary condition for the appearance of a yield stress.  

Our observations reveal that, in fact, there are two important transitions encoded in the small $f$ behavior of $P(f)$.  First, at a low packing fraction, there is the network transition where interactions between grains change from binary collisions to force networks.  This occurs in the inertial flow regime and is accompanied by a change in $P(f)$ where the peak that was present for small densities disappears and the maximum value of $P(f)$ occurs at $f=0$.  Then, as shown elsewhere, there is another transition at higher packing fraction where the system develops a yield stress and the peak reappears in $P(f)$.   

In summary, we have measured contact force distribution in this section to determine the effects of long-range correlations.  We find that $\xi/\xi_\mathrm{el} = 1.25$ separates the regime where only binary collisions occur from the regime where force networks form.   This observation allows us to split the inertial regime where hard-sphere granular flows exist into two distinct regions.  At low packing fraction there is a ``dilute regime'' where binary collisions are the dominant microscopic interaction and $\xi/\xi_\mathrm{el} < 1.25$.  At high packing fraction there is a ``dense regime'' where force networks exist but do not percolate through the system.  This dense regime is characterized by clusters of interacting grains with an average extent $\xi/\xi_\mathrm{el}$ in the range of $1.25 < \xi/\xi_\mathrm{el} < \infty$.
For dilute flows where only binary collisions occur, a peak is visible in $P(f)$; as force networks begin to appear in the dense regime, the peak disappears.  

\begin{figure}[htbp]
\begin{center}
\psfrag{yl}{\Huge{$\nu_\mathrm{bc}$}}
\psfrag{xl}{\Huge{$e$}}
\scalebox{0.37}{\includegraphics{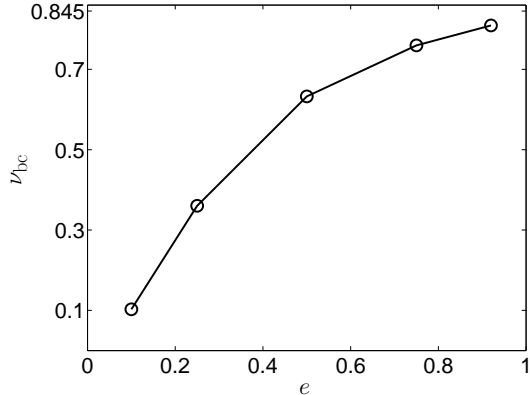}}
\caption{\label{nubce} The value of $\nu_\mathrm{bc}$ as a function of restitution coefficient $e$.  Below $\nu_\mathrm{bc}(e)$ only binary collisions are relevant and above $\nu_\mathrm{bc}$ force networks emerge.  As $e$ approaches unity the network transition is concurrent with the jamming transition at $\nu_\mathrm{c} \approx 0.845$.    
}
\end{center}
\end{figure}

The crossover at $\xi/\xi_\mathrm{el} = 1.25$ defines the transition between interactions dominated by binary collisions and interactions dominated by force networks.  Therefore, we use this value to define $\nu_\mathrm{bc}(e)$, which is the value of the packing fraction below which only binary collisions are relevant (shown schematically in Figure~\ref{phasediagram0}).  This function is plotted in Figure~\ref{nubce} using the data for $\xi$ from Figure~\ref{figxi}.  This plot is for frictionless materials $\mu=0$, but increasing the value of $\mu$ does not change the curve.  This is because, as we saw in Figure~\ref{figmuxi}, the effects of friction do not take hold until $\xi$ is much larger than $1.25 \xi_\mathrm{bc}$.  Thus $\nu_\mathrm{bc}$ is an important packing fraction for all values of friction and we see that it is an increasing function of $e$.  This is because larger $e$ produces less energy dissipation and restricts grain clustering.  The data in Figure~\ref{nubce} implies that as $e \rightarrow 1$ then $\nu_\mathrm{bc} \rightarrow \nu_\mathrm{c}$, which is the packing fraction at which the system jams.  This means that as grains become perfectly elastic and no energy is dissipated at contacts, then binary collisions describe the interactions for all values of packing fraction in the inertial regime.

\section{Conclusion} 
\begin{figure}
\psfrag{xl}{\Huge{$\nu$}}
\psfrag{yl}{\Huge{$e$}}
\psfrag{yodog}{\Huge{$\quad \nu_\mathrm{c}$}}
\resizebox{!}{.38\textwidth}{{\includegraphics{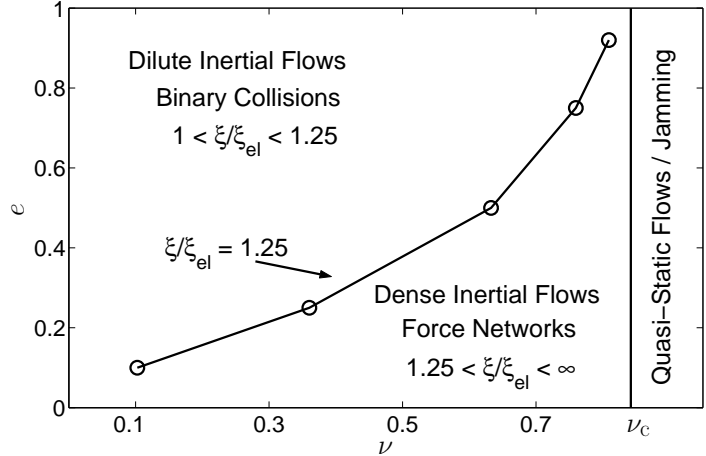}}}
\caption{\label{phasediagram} State diagram of granular shear flow, plotted as a function of the packing fraction $\nu$ and the restitution coefficient $e$.  For our system, we estimate that $\nu_\mathrm{c} \approx 0.845$, which is represented by the vertical line on the plot.  The value of $\nu_\mathrm{c}$ separates the inertial regime from the quasi-static regime.  In this paper we have focused on the inertial regime and shown that there is an important packing fraction, $\nu_\mathrm{bc}$, where $\xi/\xi_\mathrm{el} = 1.25$.  Our measured values of $\nu_\mathrm{bc}$ depend on the restitution coefficient and are plotted on the diagram.  For $\nu < \nu_\mathrm{bc}$ the flow is in the Dilute Inertial Regime where microscopic interactions consist of binary collisions; for $\nu > \nu_\mathrm{bc}$ the flow is in the Dense Inertial Regime where long-ranged force networks begin to form.
}
\end{figure}

We have measured correlations between grain-forces in inertial shear flows.  These correlations are long-ranged, decaying with a characteristic length scale $\xi$ that diverges at the jamming transition and asymptotes to an elastic value $\xi_\mathrm{el}$ in the dilute limit.  By investigating the statistics of contact forces between grains, we have shown that $\xi/\xi_\mathrm{el} = 1.25$ splits the inertial regime into dilute flows, where all forces arise from binary collisions between grains, and dense flows, where force chain networks begin to form.  We denoted $\nu_\mathrm{bc}$ as the packing fraction at which $\xi/\xi_\mathrm{el} = 1.25$ and found that the value of $\nu_\mathrm{bc}$ depends on the restitution coefficient, but is always less than the packing fraction $\nu_c$ at which the system jams.  This phenomenology is illustrated in Figure~\ref{phasediagram}.

The crossover from the dilute to dense regime is accompanied by a qualitative change in the nature of contact forces between grains, measured using the contact force distribution function $P(f)$.  For shear flows in the dilute regime $P(f)$ has a peak at non-zero $f$, whereas for shear flows in the dense regime there is no peak.  This raises interesting questions about the nature of the transition that occurs at $\nu_\mathrm{bc}$.  We observe that the average coordination number $z$ is equal to one for flows with $\nu < \nu_\mathrm{bc}$ and is greater than one for $\nu>\nu_\mathrm{bc}$.  Further investigation of the behavior of granular flows very close to $\nu_\mathrm{bc}$ is needed to bring further insight to the physics near the transition.

From a wider viewpoint, the presence of long range correlations changes the assumptions that can be made when modeling properties of the system.  We explore the theoretical implications of correlations in the companion paper~\cite{gajcompanion2}. 

This work was supported by the William M. Keck Foundation, the MRSEC program of NSF under Award No. DMR00-80034, the James S. McDonnell Foundation, the David and Lucile Packard Foundation, and NSF Grant Nos. DMR-9813752, PHY99-07949 and DMR-0606092.

\end{document}